%% file: hybrid-bc.tex
\documentclass[conference,10pt]{IEEEtran}

\IEEEoverridecommandlockouts

\usepackage{url}
\usepackage{cite}
\usepackage{graphicx}
\usepackage{amsmath}
\usepackage{subfigure}
\usepackage{color}
\usepackage{caption}
\usepackage[ruled,figure,linesnumbered]{algorithm2e}
\usepackage{comment}

\newcommand{\REM}[1]{}
\newcounter{todocounter}

\begin{document}

\title{A Fine-Grained Hybrid CPU-GPU Algorithm for Betweenness Centrality Computations}

\author{\IEEEauthorblockN{$^1$Ashirbad Mishra, $^1$Sathish Vadhiyar, $^2$Rupesh Nasre, $^{3,4}$Keshav Pingali}
\IEEEauthorblockA{
$^1$Supercomputer Education and Research Centre, Indian Institute of Science, Bangalore, India \\
$^2$Department of Computer Science and Engineering, Indian Institute of Technology, Madras, India \\
$^3$Institute for Computational Engineering and Sciences, University of Texas at Austin, USA \\
$^4$Department of Computer Science, University of Texas at Austin, USA \\
ashirbadm@ssl.serc.iisc.in, vss@serc.iisc.in, rupesh@cse.iitm.ac.in, pingali@cs.utexas.edu
}
}

\maketitle

\input{abstract.tex}

\input{intro.tex}

\input{background.tex}
\input{related.tex}

\input{distance_cal.tex}

\input{algorithm_implementation_variablepart.tex}

\input{exp_res.tex}
\input{confut.tex}

\bibliographystyle{IEEEtran}
\begin{footnotesize}
\bibliography{hybrid-bc}
\end{footnotesize}

\end{document}

%% file: abstract.tex
\begin{abstract}
 
Betweenness centrality (BC) is an important graph analytical application for large-scale graphs. While there are many efforts for parallelizing betweenness centrality algorithms on multi-core CPUs and many-core GPUs, in this work, we propose a novel fine-grained CPU-GPU hybrid algorithm that partitions a graph into CPU and GPU partitions, and performs BC computations for the graph on both the CPU and GPU resources simultaneously with very small number of CPU-GPU communications. The forward phase in our hybrid BC algorithm leverages the multi-source property inherent in the BC problem. We also perform a novel hybrid and asynchronous backward phase that performs minimal CPU-GPU synchronizations. Evaluations using a large number of graphs with different characteristics show that our hybrid approach gives 80\% improvement in performance, and 80-90\% less CPU-GPU communications than an existing hybrid algorithm based on the popular Bulk Synchronous Paradigm (BSP) approach.
\end{abstract}

%% file: intro.tex
\section{introduction}

Large scale network analysis is prevalent in diverse networks such as social, transportation and biological networks. In most network analysis, graph abstractions and algorithms are frequently used to extract interesting information. Real world networks are often very large in size resulting in graphs with several hundreds of thousands to billions of vertices and edges. Recently, GPUs have been used successfully in accelerating different graph algorithms \cite{merrill-scalableGPU-ppopp2012,gharaibeh-blindDating-ipdps2013,mclaughlin-scalableBCGPU-sc2014,sariyuce-regularizingCentrality-jpdc2015,slota-HPGraphAnalytics-ipdps2015}.
Centrality metrics, such as betweenness and closeness quantify how central a node is in a network. They have been used successfully for various analyses including quantifying importance in social networks \cite{merrer-socialGraphs-sns2009}, studying structures in biological networks \cite{delsol-topologyProtein-bio2005}, analysis of covert network \cite{coffman-intelligenceAnalysis-CommACM2004}, and for identifying hubs in transportation networks \cite{guimera-airNetwork-NAS2005}.

Betweenness centrality finds the node through which the maximum number of shortest paths pass. The algorithm by Brandes \cite{brandes-bc-jms2001} forms the basis of most of the work on between centrality. For each node of the graph, the algorithm consists of a forward phase that finds the shortest path between the node as a source and the other nodes using BFS or SSSP, and a backward phase that computes {\em dependency scores} of the non-source nodes. These dependency scores are summed across shortest paths with different sources nodes to compute centrality scores of the nodes. The earlier efforts on parallel computations of betweenness centrality were primarily on multi-core CPUs \cite{bader-centralityIndices-icpp2006,madduri-multithreadedBC-ipdps2009,prountzos-BC-ppopp2013}. Subsequently, many of the recent efforts have been on many-core GPUs \cite{jia-edgevnodeCentrality-gpugems2011,shi-fastnetworkcentralityGPU-bmcbio2011,mclaughlin-scalableBCGPU-sc2014}. The GPU-only strategies, while providing high performance for small graphs, are limited in terms of exploring large graphs due to the limited memory available on GPU. They also do not utilize the power of host multi-core CPUs that are typically connected to the GPU devices. A hybrid strategy involving computations on both the CPU and GPU cores can help explore large graphs and utilize all the resources. Some of the GPU-based strategies adopt coarse-level hybridization for betweenness centrality in which the CPU and GPU perform the entire betweenness centrality algorithm, but for different sources \cite{sariyuce-bc-gpgpu2013}. However, this can result in sub-optimal work distribution to the CPU and GPU cores and hence idling of the resources due to different performance of the CPU and GPU and the different workloads for different sources. Fine-level hybridization partitions a graph and performs computations for a single source on both the CPU and GPU partitions. Such fine-level hybridization also allows to explore large graphs that cannot be accommodated in either of the CPU or GPU memory units but can be accommodated in the combined memory size. Totem \cite{gharaibeh-oxenChicken-pact2012}, a framework for fine-level hybridization, adopts level-wise BFS in the forward phase across both the CPU and GPU cores, resulting in a large number of communications and synchronizations between the CPU and GPU. Moreover, the existing efforts on betweenness centrality primarily focus on optimizing the BFS or SSSP in the time-consuming forward phase for a single source, and not necessarily considering the property of betweenness centrality problem that involves computations for a large number of sources. 

Thus, a fundamental rethink of the algorithmic steps is required in performing fine-level hybridization to minimize resource idling while avoiding excess synchronization and communications between the CPU and GPU, and leveraging the multi-source property inherent in the betweenness centrality problem. In this paper, we propose a novel fine-grained CPU-GPU hybridization strategy in which we partition the graph into CPU and GPU partitions, and formulate the betweenness centrality in terms of the distances between the border nodes in each partition that are computed independently on the CPU and GPU using only the nodes and edges of the respective partitions. These distances are stored in {\em border matrices}, one each for the CPU and GPU partitions. The one-time computation of border matrices is then harnessed for the betweenness centrality computations in multiple source nodes, where for each source node, our algorithm performs an {\em iterative refinement} of the border node distances from the source, followed by {\em simultaneous relaxation} of the distances by the CPU and GPU in their respective partitions for the forward phase. We also perform a novel hybrid and asynchronous backward phase that performs maximum amount of independent computations and minimal CPU-GPU synchronizations and communications. Comparisons with an existing hybrid algorithm based on the popular Bulk Synchronous Paradigm (BSP) approach show about 80\% improvement in performance, and 80-90\% less CPU-GPU coordinations with our approach.

Section \ref{back} gives the background related to BC computations, and Section \ref{related} surveys the existing work on parallel BC. Section \ref{distance} gives our hybrid algorithm for calculating distances of the nodes in the shortest paths using both the CPU and GPU, along with the proofs for correctness and convergence. Section \ref{algorithm} gives the rest of the details of the algorithm including $\sigma$ computations, our novel backward phase algorithm and implementation details. Section \ref{exp_res} presents experiments and results comparing our approach with Totem. Finally, Section \ref{con_fut} gives conclusions and future work.

%% file: background.tex
\section{Background}
\label{back}

Let $G=(V,E)$ be a graph with $n$ vertices and $m$ edges. Betweenness centrality of a vertex $v$, $BC[v]$, is defined as:
\begin{equation}
\label{bc-eq-1}
 BC[v] = \sum_{s \neq v \neq t \in V} \frac{\sigma_{st(v)}}{\sigma_{st}}
\end{equation}
where $\sigma_{st}$ is the number of shortest paths between two vertices, $s$ and $t$, and $\sigma_{st(v)}$ is the number of those shortest paths passing through $v$. The fraction in Equation \ref{bc-eq-1} is denoted as $\delta_{st}(v)$, the {\em pair dependency} of $v$ on the pair $s$ and $t$. One way of calculating $BC[v]$ is to find the shortest paths between all pairs and keep track of the number of shortest paths passing through $v$. However, this involves a complexity of $O(n^3)$.

Brandes \cite{brandes-bc-jms2001} proposed an algorithm in which the pair dependencies are accumulated over all the target vertices to form {\em source dependency} of $v$ on a source $s$, $\delta_s(v) = \sum_{t \neq v} \delta_{st}(v)$. This {\em source dependency}, $\delta_s(v)$, is calculated using a recursive formulation:
\begin{equation}
 \label{bc-eq-2}
 \delta_s(v) = \sum_{u: v \in P_s(u)} \frac{\sigma_{sv}}{\sigma_{su}}(1+\delta_s(u))
\end{equation}
$P_s(u)$ is the set of predecessors of $u$ in the shortest paths from $s$. The betweenness centrality, $BC[v]$, is then given by $BC[v] = \sum_{s \neq v \in V}\delta_s(v)$. The algorithm consists of two phases: a {\em forward phase} and a {\em backward phase}. The forward phase consists of a BFS traversal or SSSP calculation with $s$ as the source. For every vertex, $v$, visited in the forward phase, the distance of the vertex from the source, the number of shortest paths through $v$, $\sigma_{sv}$ and the set of predecessors are calculated and updated. The {\em backward phase} traverses the vertices in the descending order of the distances from the source to compute $\delta_s(v)$ of a predecessor, $v$, using the $\delta$ scores of its successors that are already computed.
The total complexity of BC with Brandes' algorithm is thus $O(mn)$ corresponding to $m$ BFS traversals for every source.

The algorithm for sequential betweenness centrality for a source, $s$, is shown in Figure \ref{seq-BC}. To parallelize the forward phase, the for loops in lines \ref{node-level-for1} and \ref{edge-level-for1} can be performed in parallel by multiple threads. Parallelizing only the outer-level for loop in line \ref{node-level-for1} will amount to {\em vertex-parallel} algorithm, while parallelizing both the for loops will amount to {\em edge-level} parallelism. Simultaneous access to common data structures, namely, $dist$, $sigma$, and $pred$, by multiple threads in the for loops will have to be protected by atomic constructs or locks. Similarly, the backward phase is parallelized by performing the for loops in lines  \ref{node-level-for2} and \ref{edge-level-for2} in parallel. The computations can also be organized as {\em topology-driven} or {\em data-driven}. In topology-driven approach, all the vertices or edges of the graph are assigned to the threads, and at a given time step only those threads owning vertices/edges that have to be processed in the time step (a.k.a. {\em active elements}) perform computations. In data-driven approach, a dynamic worklist maintains only the active elements for a time step and the threads are assigned only to these active elements.

\begin{algorithm}[t]
\begin{footnotesize}
\SetKwInOut{Input}{input}
\SetKwInOut{Output}{output}
\SetKwProg{Fn}{}{ \{}{\}}

\Input{a graph $graph(N,M)$ with $N$ vertices in a set $V$, and $M$ edges in a set, $E$. A source $s$}


    $level[0] \leftarrow {s}$; $dist[s] \leftarrow 0$; $\sigma[s] \leftarrow 1$; \\
    $dist[\forall v \in V \backslash {s}] \leftarrow -1$; $\sigma[\forall v \in V \backslash {s}] \leftarrow 0$; $pred[\forall v \in V \backslash {s}] \leftarrow \emptyset;$ \\
    \BlankLine

    $curLevel \leftarrow 0;$ \\
    \tcc{Forward Phase}
    \While{$level[curLevel] \neq \emptyset$}{
      \For{$v \in level[curLevel]$}{
\label{node-level-for1}
        \For{$w \in neighbors(v)$}{
\label{edge-level-for1}
          \If{$dist[w] = -1$}{
            $level[curLevel+1] \leftarrow level[curLevel+1] \cup w$;
            $dist[w] \leftarrow dist[v]+1$;
          }
          \If{$dist[w] = dist[v]+1$}{
            $\sigma[w] \leftarrow \sigma[w]+\sigma[v]$;
            $pred[w] \leftarrow pred[w] \cup v$;
          }
        }
      }
      $curLevel++$;
    }
    \BlankLine

    \tcc{Backward Phase}
    $curLevel \leftarrow curLevel-1$;
    $\delta[\forall v \in V] \leftarrow 0$; \\
    \While{$curLevel > 0$}{
      \For{$u \in level[curLevel]$}{
\label{node-level-for2}
        \ForAll{$v \in pred[u]$}{
\label{edge-level-for2}
          $\delta[v] \leftarrow \delta[v] + \frac{\sigma[v]}{\sigma[u]}(1+\delta[u])$;
        }
      }
      $curLevel--$;
    }

\caption{\small{Betweenness Centrality Algorithm}}
\label{seq-BC}
\end{footnotesize}
\end{algorithm}

%% file: related.tex
\section{Related Work}
\label{related}

There has been a number of efforts on implementing betweenness centrality on multi-core CPUs.
Bader and Madduri \cite{bader-centralityIndices-icpp2006} developed the first optimized parallel algorithms for different centrality indices including betweenness centrality on shared memory multiprocessors and multithreaded architectures.
Madduri et al. \cite{madduri-multithreadedBC-ipdps2009} proposed a lock-free fast parallel algorithm for betweenness centrality for multi-core architectures by adopting an alternate representation for predecessor multisets using a cache-friendly successor multisets.
Galois \cite{pingali-tao-pldi2011} is a system for multi-core environments that incorporates the concept of the operator formulation model in which an algorithm is expressed in terms of its action (or operator) on data structures. It has been used to provide large-scale performance for many graph based algorithms including betweenness centrality \cite{prountzos-BC-ppopp2013}.
All these efforts on multi-core CPUs can potentially gain in performance by including GPU computations on heterogeneous systems.

There have been recent efforts on accelerating BC computations on GPUs \cite{jia-edgevnodeCentrality-gpugems2011,shi-fastnetworkcentralityGPU-bmcbio2011,sariyuce-bc-gpgpu2013,mclaughlin-scalableBCGPU-sc2014}.
In a recent work, McLaughlin and Bader \cite{mclaughlin-scalableBCGPU-sc2014} have developed scalable BC computations for GPUs. Their strategies include a work-efficient parallel algorithm that employs vertex-based parallelism, explicit queues for graph traversal, compact data structures by discarding the predecessor array, CSR data structure for distinguishing levels in the dependency accumulation stage and utilizing different blocks on different SMs to process multiple roots.
While the GPU-only solutions can potentially increase performance, they are limited by the sizes of the graphs that can be processed due to limited GPU memory.

The work by Sariy{\"{u}}ce et al. \cite{sariyuce-bc-gpgpu2013} is one of the first efforts that explored hybrid computations utilizing both the CPU and GPU cores.
They perform coarse-grain hybrid parallelism on heterogeneous CPU-GPU architectures by processing independent BC computations on different roots simultaneously on CPUs and GPUs. In their recent work \cite{sariyuce-regularizingCentrality-jpdc2015}, they partition the GPU threads among multiple simultaneous BFS traversals for BC computations of multiple sources, similar to the work by McLaughlin and Bader \cite{mclaughlin-scalableBCGPU-sc2014}. In their scheme, a set of consecutive threads process a virtual vertex for multiple BFSs for different sources, thereby employing interleaved BFSs.

Fine-level hybridization can help explore large graphs that can fit only within the combined memory size of the CPU and GPU. Totem \cite{gharaibeh-oxenChicken-pact2012} is a graph processing engine that partitions \cite{gharaibeh-blindDating-ipdps2013} the graph across the CPU and GPU of a heterogeneous system and processes the fine-level computations simultaneously on both the CPU and GPU cores. The Totem programming model follows Bulk Synchronous Programming (BSP) model which involves communication and synchronization between the CPU and GPU for each superstep. This results in CPU-GPU communication for every level in the BFS forward phase of the BC computations, while in our work, the number of CPU-GPU communications is related to the number of iterations for convergence, which in most cases have been found to be less than ten.

%% file: distance_cal.tex
\section{Hybrid CPU-GPU Distance Calculations}
\label{distance}

In our hybrid algorithm, the given graph G is partitioned into a CPU partition and a GPU partition. {\em Border edge} is an edge that has one end point in one partition and the other end point in another partition. The end point of a border edge is called as a {\em border node}. The hybrid algorithm has the following steps.

\subsection{Notations}

\noindent
1. $P_{C}$, $P_{G}$: The CPU and GPU partitions of the graph, respectively. \\
2. $B_{G}$: set of all border nodes in the graph. \\
3. $B_{P_{C}}$, $B_{P_{G}}$: Border nodes in CPU and GPU partitions, respectively. \\
4. $Pr(u)$: The partition of the graph $G$ to which the vertex $u$ belongs. \\
5. $B_{Pr(u)}$: The set of border nodes in the partition to which the vertex $u$ belongs. \\
6. $d_{C}[u,v]$ : The shortest path distance from vertex $u$ to vertex $v$ computed by our hybrid algorithm.


\subsection{Border Matrix Computations}

This step is a preprocessing step which is performed only once for the entire graph G. In the CPU partition, considering each border node $b_{i}$ as source at a time, a BFS/SSSP is performed which computes $d_{C}[b_{i},v]$ $b_{i}\in B_{P_{C}}\: \wedge \forall v\in P_{C}$. The result is a border matrix $BM_{P_{C}}$ which stores the distance value of the shortest path between each pair of border nodes in the partition $P_{C}$. i.e 
\begin{equation}
BM_{P_{c}}[i][j] = d_{C}[b_{i},b_{j}]
\end{equation}
$\forall i,j$ where $b_{i},b_{j} \in B_{P_{C}}$. Similarly, $BM_{P_{G}}[i][j]$ is computed for the GPU partition. Both $BM_{P_{C}}$ and $BM_{P_{G}}[i][j]$ are computed in parallel and asynchronously on the CPU and GPU.

\subsection{Distance Calculations in the Forward Phase for a source}

A source $s$ is selected on which BFS/SSSP is to be performed. All the nodes in the graph $G$, except $s$ are initialized to $\infty$. $s$ is initialized to 0. Our hybrid algorithm performs the distance calculations as follows:

\noindent
{\bf Step 1:} BFS/SSSP from source $s$ in the partition Pr(s). This step computes $d_{C}[s,v]$, $\forall v \in Pr(s)$. We also denote this step as the {\em initial BFS/SSSP} step.

\noindent
{\bf Iterations of steps 2-5:}

The computations in Step 1 results in a set of distance values for border nodes. i.e $d_{C}[s,b_{i}]$, $\forall b_{i} \in B_{Pr(s)}$. The following steps are iterated until the termination condition is satisfied.

\noindent
{\bf Step 2: Updates of $B_{G-Pr(s)}$ using edge cuts}

The distance values of the border nodes in the partition $G-Pr(s)$, i.e., in the non-source partition, are updated using all the edge cuts or edges that connect the border nodes of two partitions. The distance values of vertices in $B_{G-Pr(s)}$ are updated as follows: \\
$\forall b_{i} \in B_{Pr(s)} , b_{j} \in B_{G-Pr(s)}$ \\
if $d_{C}[s,b_{j}] \geq d_{C}[s,b_{i}]+w(b_{i},b_{j})$, then\\
$d_{C}[s,b_{j}]=d_{C}[s,b_{i}]+w(b_{i},b_{j})$, where $w(b_{i},b_{j})$ is the weight of the edge cut, $e(b_{i},b_{j})$.

\noindent
{\bf Step 3: Updates of $B_{G-Pr(s)}$ using border matrix, $BM_{G-Pr(s)}$}

In this step, the distance values of vertices in $B_{G-Pr(s)}$ from $s$ are refined from the earlier computed values using the border matrix, $BM_{G-Pr(s)}$, as follows: \\
$\forall b_{i},b_{j} \in B_{G-Pr(s)}$ \\
if $d_{C}[s,b_{j}] \geq d_{C}[s,b_{i}]+BM_{G-Pr(s)}[i][j]$, then\\
$d_{C}[s,b_{j}]=d_{C}[s,b_{i}]+BM_{G-Pr(s)}[i][j]$

\noindent
{\bf Step 4: Updates of $B_{Pr(s)}$ using edge cuts}

This step is similar to Step 2, but is used to update the distances of the border nodes in $Pr(s)$, $B_{Pr(s)}$, using the distances of the border nodes in $G-Pr(s)$, $B_{G-Pr(s)}$, and the weights of the edge cuts.

\noindent
{\bf Step 5: Updates of $B_{Pr(s)}$ using border matrix, $BM_{Pr(s)}$}

This step is similar to step 3, but is used to update the distances of the border nodes, $B_{Pr(s)}$, using the border matrix, $BM_{Pr(s)}$.

Steps 2-5 are iterated for multiple times until the distance values of the border nodes in $B_{Pr(s)}$, before and after step 5 are the same. Of these, steps 2 and 4 require CPU-GPU communications, while steps 3 and 5 are performed independently.

\noindent
{\bf Step 6: Edge relaxation for finding the final distances of non-border nodes}

After the termination of the iterations, step 2 is performed once so that the distance values of $B_{G-Pr(s)}$ are updated correctly. Then, CPU and GPU relax the edges in their own partition, $P_{C}$ and $P_{G}$, respectively, computing the following:\\
$d_{C}[b_{i},v]$~$(\forall b_{i}\in B_{G}) \wedge (\forall v \in G) \wedge (Pr(b_i) = Pr(v)$), \\
using the distance values of all the border nodes, $B_{G}$. We denote this step as the {\em relaxation} step.

\subsection{Proof of Correctness}

\begin{figure}
 \centering
 \includegraphics[scale=0.8]{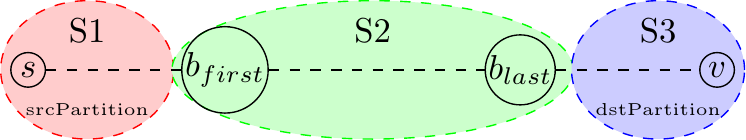}
 \caption{\small{Decomposition of a shortest path between $s$ and $v$}}
 \label{shortest-decomp}
\end{figure}

Consider a shortest path from root node $s$ to terminal node $v$ cutting across the partitions multiple times. Let the partition containing $s$ be denoted as $srcPartition$ and the partition containing $v$ referred as $dstPartition$. Such a shortest path can be decomposed into three sets: \\
1. set $S1$, which is the starting sequence in the shortest path starting at the root node $s$, containing nodes and edges only belonging to $srcPartition$, and ending at a border node $b_{first}$ in the $srcParition$ such that the next edge in the shortest path after the set $S1$ is an edge cut connecting $b_{first}$ to a node in the non-$srcPartition$, \\
2. set $S2$ containing the intermediate edges of the path spanning the partitions, and \\
3. set $S3$, which is the ending sequence in the shortest path starting at a border node $b_{last}$ in the $dstPartition$, containing nodes and edges only belonging to $dstPartition$, and ending at the terminal node $v$, such that the previous edge in the shortest path before $S3$ is an edge cut connecting a node in the non-$dstPartition$ and $b_{last}$. \\
The decomposition of the shortest path is shown in Figure \ref{shortest-decomp}.

\begin {figure}
\centering
\subfigure[Case 1]{
\includegraphics[scale=0.8]{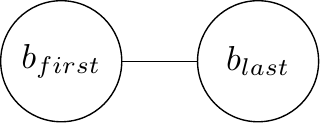}
\label{s2-case1}
}
\subfigure[Case 2] {
\includegraphics[scale=0.8]{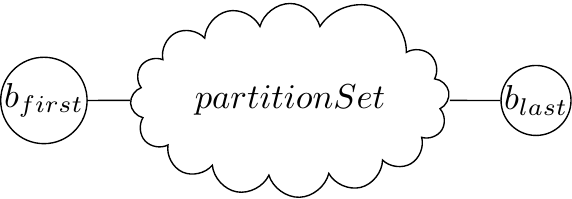}
\label{s2-case2}
}
\subfigure[Case 3] {
\includegraphics[scale=0.8]{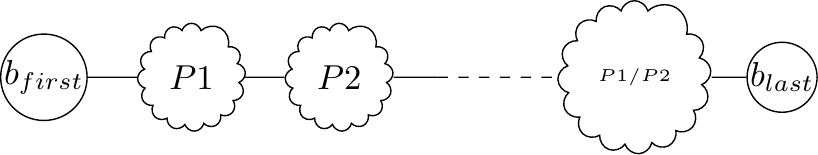}
\label{s2-case3}
}
\caption{\small{Three Cases in S2}}
\label{s2-cases}
\end{figure}

$S2$ contains the intermediate path starting with an edge cut with source node $b_{first}$ and ending with an edge cut whose destination node is $b_{last}$. Note that $S2$ can contain only a single edge cut with source and destination nodes as $b_{first}$ and $b_{last}$, respectively. If $S2$ contains multiple edge cuts, two consecutive edge cuts in $S2$ are connected by zero or more edges belonging to a single partition, denoted as $partitionSet$. $S2$ can contain multiple such $partitionSets$ with two consecutive $parititonSets$ corresponding to two different partitions, $P1$ and $P2$, respectively, and the edge cuts after them connecting $P1$ to $P2$, and $P2$ to $P1$, respectively. These three cases in S2 are illustrated in Figure \ref{s2-cases}.

To prove the correctness of the algorithm, we need to show that this shortest path from $s$ to $v$ can be determined by our algorithm.

\noindent
{\bf Set S1:}

Our algorithm, in step 1, finds the distances of the border nodes in $srcPartition$. At least one border node will get its final correct distance, i.e., the shortest distance from $s$, in this step (vide proof of convergence below). By the definition of $b_{first}$, it is one of the border nodes that will get its final correct distance from the root node $s$ in step 1. On the contrary, if $b_{first}$ does not get its correct distance in this step, then its distance will be corrected in the subsequent steps, implying that the shortest path to $b_{first}$ is through an edge cut involving another border node in $srcParition$, contradicting the definition of $b_{first}$.

\noindent
{\bf Set S2:}

Multiple border nodes in the non-$srcPartition$ can be directly connected to $b_{first}$ using edge cuts. Step 2 of our algorithm finds the distances to these border nodes using the edge cut weights. At least one of these distances will be the final correct distance. The border node that is connected to $b_{first}$ in the $s-v$ shortest path will get its correct distance from $s$ in step 2 of our algorithm. On the contrary, if the distance to this border node is updated in the subsequent steps, then our shortest path will contain some other edge cut from $b_{first}$ to some other node, contradicting the shortest path claimed.

If this border node, $b1_{nonsrc} \ne b_{last}$, then the path from $b1_{nonsrc}$ has to traverse back to the $srcPartition$. There are two possibilities: \\ \\
\begin {figure}
\centering
\subfigure[Case 1]{
\includegraphics[scale=0.6]{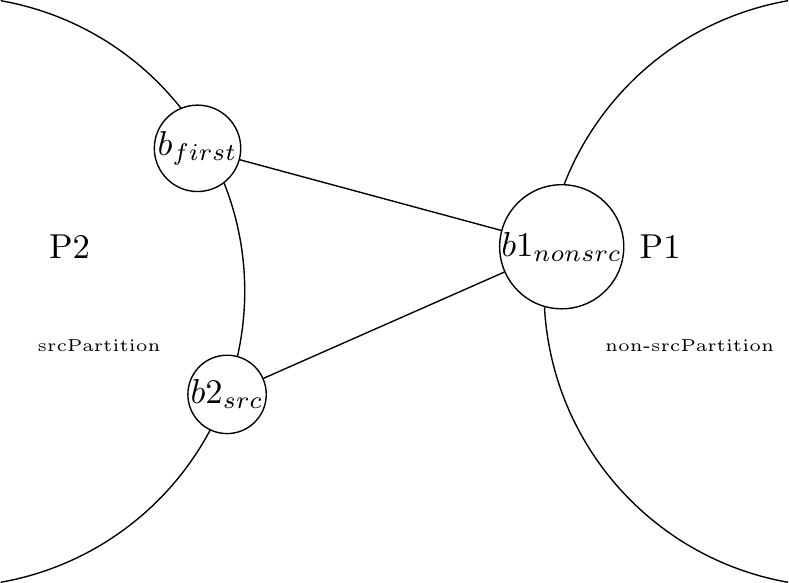}
\label{s2-border-case1}
}
\subfigure[Case 2] {
\includegraphics[scale=0.6]{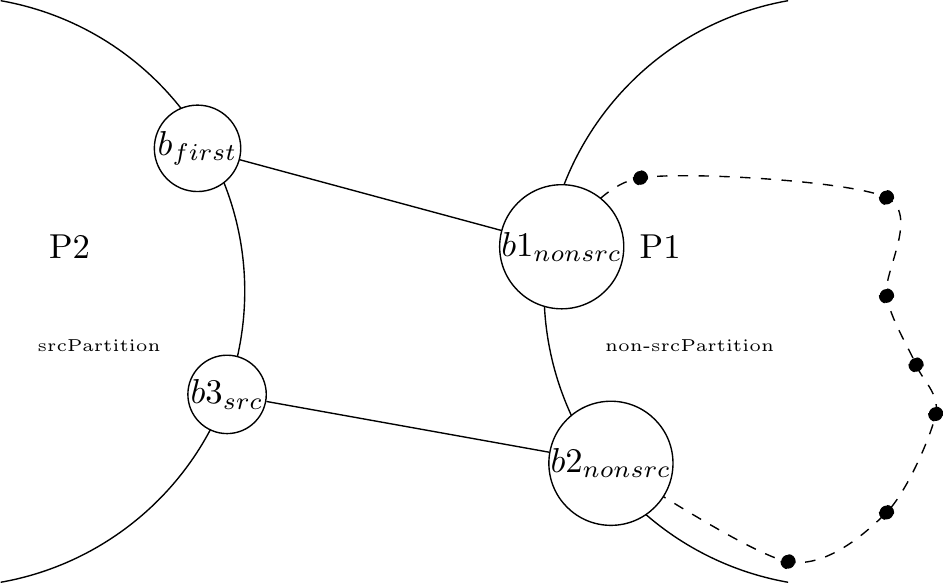}
\label{s2-border-case2}
}
\caption{\small{Cases with Border Nodes in S2}}
\label{s2-border-cases}
\end{figure}
1. The next edge from $b1_{nonsrc}$ in our shortest path from $s$ to $v$ is an edge cut to a border node $b2_{src}$ in $srcPartition$ as shown in Figure \ref{s2-border-case1}. In this case, we need to prove that the distance to $b2_{src}$ from $s$ is correctly updated by our algorithm and this distance will be lesser than the distances of any other path through $b1_{nonsrc}$ that traverses the nodes and edges in the non-$srcPartition$ before reaching $b2_{src}$. Step 3 of our algorithm finds the distances, at this stage, from $s$ to the border nodes in the non-$srcPartition$ using the intermediate distances of these border nodes found in step 2 and the $BM$ matrix that denotes the paths between these border nodes that traverse only the nodes and edges in the non-$srcPpartition$. \\
a. If these distances to the border nodes other than $b1_{nonsrc}$ are all smaller than the distances from $s$ to these border nodes through $b1_{nonsrc}$, then the shortest path from $s$ to $v$ through $b1_{nonsrc}$ traverses back to the $srcPartition$ only through an edge cut to a border node in the $srcPartition$ as the next edge. Step 4 of our algorithm finds the distances to these border nodes in $srcPartition$ that are connected to $b1_{nonsrc}$ using the edge cut weights. The border node $b2_{src}$ that is connected to $b1_{nonsrc}$ in the $s-v$ shortest path will get its correct distance from $s$ in step 4 of our algorithm due to a similar reasoning as applied for $b1_{nonsrc}$ above. \\
b. On the other hand, if some of the distances to the border nodes in the non-$srcPartition$ other than $b1_{nonsrc}$, found by step 3, are equal to the distances from $s$ to these border nodes through $b1_{nonsrc}$, then our step 4 compares the distance to $b2_{src}$ from $b1_{nonsrc}$ via the edge cut with the distance to $b2_{src}$ from $b1_{nonsrc}$ via another border node in the non-$srcParition$ and chooses the smaller of these two. Since our shortest path has the edge cut to $b2_{src}$ as the next edge, this $b1_{nonsrc}-b2_{src}$ edge cut weight must be smaller than the distance from $b1_{nonsrc}$ to $b2{src}$ via another border node in the non-$srcPartition$. \\

\noindent
2. The next edge in our shortest path from $s$ to $v$ after $b1_{nonsrc}$ is an edge belonging to the non-$srcParition$, and after a succession of edges in the non-$srcParition$, the shortest path contains an edge cut to a node $b3_{src}$ in the $srcPartition$ from another border node $b2_{nonsrc}$ in the non-$srcPartition$, as shown in Figure \ref{s2-border-case2}.
 In this case, we need to prove that the distance to $b3_{src}$ is correctly updated by our algorithm and this distance through $b2_{nonsrc}$ will be lesser than the weight on the edge cut that may exist between $b1_{nonsrc}$ and $b3_{src}$. At least one of the distances from $s$ through $b1_{nonsrc}$ to one of the other border nodes in the non-$srcPartition$ through only the nodes and edges of the non-$srcPartition$, found by step 3 of our algorithm, will be the correct final distance of this border node. On the contrary if none of these distances is the correct final one, then the $s-v$ shortest path either does not pass through $b1_{nonsrc}$ at all or the next edge in the shortest path from $s$ to $v$ through $b1_{nonsrc}$ will be an edge cut from $b1_{nonsrc}$ to the $srcPartition$, contradicting the shortest path claim. The border node, $b2_{nonsrc}$ will be one of these nodes that will get its correct final distance from $s$. On the contrary, if its correct distance will be updated in the subsequent steps, then the shortest path through $b1_{nonsrc}$ will get back to the $srcPartition$ through some other border node of the other partition, contradicting the shortest path claim. By similar reasonings as above for Set S1 and point 1.b, some of the border nodes in the $srcParition$ connected by edge cuts from $b2_{nonsrc}$ will get their correct distances by step 4, $b3_{src}$ will be one of these border nodes, and this correct distance through $b2_{nonsrc}$ will be smaller than the edge cut weight that may exist between $b1_{nonsrc}$ and $b3_{src}$ during the comparisons made by step 4 of our algorithm.

If the border nodes, $b2_{src}$ and $b3_{src}$ are not equal to $b_{last}$, then the path from these border nodes will have to traverse back to the non-$srcPartition$. The same arguments used above are extended for the $srcParition$, this time using steps 4 and 5. Thus our algorithm progressively finds the correct distances of the nodes in the set S2 of our $s-v$ shortest path in the increasing order of path lengths, traversing back and forth between the source and the other partitions until it reaches the border node, $b_{last}$, in the destination partition.

\noindent
{\bf Set S3:}

Having found the correct distance to $b_{last}$, our algorithm continues with the iterations of the steps 2-5 until it finds the correct distances of all the other border nodes in the destination partition containing $v$. From $b_{last}$, it finds the correct distance to $v$ using only the destination nodes and edges using the standard BFS/SSSP procedure, and hence the proof of this is trivial.

\subsection{Proof of Convergence}

From the above proof of correctness, we find that every time when a shortest path traverses from one partition to another, the distances of two border nodes, one each in a partition and connected by an edge cut, converge to their final correct distances. Each iteration spanning steps 2 to 5 of our algorithm traverses from a border node in one partition to another, back to a border node in the first partition either directly or through another border node in the second partition, thus converging to the final distances of two border nodes, one in each partition in the worst case. Thus, in the worst case, when the longest of the shortest path between $s$ and another node traverses across the partitions multiple times and passes through each of the border nodes, the total number of iterations is equal to the maximum of the number of border nodes in both the partitions.

\subsection{Space Complexity}

We consider equal-size partitions in the partitioned case. We consider a graph $G(V, E)$ with with a set of vertices $V$ of size $n$ and set of edges $E$ of size $m$.

\subsubsection{Single-device Algorithm with Unpartitioned Graph}

The graph is stored in a hybrid CSR-COO format. The format contains an offset array of size $n$, which stores the offset to the adjacency array for each vertex. The source and destination of each edge is stored in separate arrays, each of size $m$. An additional array of size $m$ is used to store the weights of the edges. Thus the total size for these arrays is $3m+n$. Besides, we use three arrays to store the meta-data of each vertex of the graph. One array stores the distance from source, the second stores the sigma values and the third stores the delta values of each vertex. The total size of these vertex-based arrays is $3n$. Finally, an array of size $n$ is use to store the sigma value of each edge of the graph. Hence, the total space complexity for the unpartitioned case is:
\begin{equation}
\label{space_unpart}
 Space_{unpartitioned} = 4(m+n)
\end{equation}

\subsubsection{Hybrid CPU-GPU Algorithm with Equal-sized Partitions}

In the hybrid algorithm, each device consumes half the space for the arrays mentioned for the unpartitioned case, described above. In addition, the hybrid algorithm also stores information for the border nodes. Considering equal number of border nodes in both the partitions, and is equal to $b$, each partition requires four arrays of size $b$ to store the information regarding their border vertices. One array stores the identity, the second stores the distance values, the third stores sigma and finally the fourth stores delta values of all border nodes in that partition. The first three arrays are used during the forward phase in the iterative step while all of the arrays are used during the backward phase for communication. Finally, each partition stores a border matrix of size $b^2$. Thus, the total space complexity of the hybrid algorithm is:
\begin{equation}
\label{space_part}
 Space_{hybrid} = 2(m+n)++4b+b^2
\end{equation}

\subsubsection{Case Study}

Consider the graph, \textit{nlpkkt220}, with the number of vertices $n=27093600$ and number of edges $m=514179537$. Substituting in Equation \ref{space_unpart}, we find that the memory requirement when the graph is unpartitioned and given as a whole to a GPU or a multi-core CPU corresponds to $2165092548$ elements. With each element being an integer, this requires 9 GB of memory. In addition to the calculated space, miscellaneous data structures (e.g., BFS queue) are also required. Current commercial GPUs (e.g., K20m) cannot accommodate such amount of space.
When the graph is partitioned using METIS, the number of border nodes, $b$, for this graph is $40$. Substituting in Equation \ref{space_part}, the memory requirement in CPU or GPU for the hybrid algorithm corresponds to $1082548034$ or about 4.5 GB, which can be accommodated in the current GPU architectures. Thus, the hybrid algorithm enables the use of GPU for the exploration of graphs that cannot entirely be accommodated in the GPU memory.

%% file: algorithm_implementation_variablepart.tex
\section{Complete Algorithm, Practical Implementation and Variable Partitioning}
\label{algorithm}

\subsection{$\sigma$ Computations and Forward Phase Algorithm}

While the previous section focused on distance computations, in this section we explain the computations of $\sigma$ values and outline the entire forward phase algorithm in general. Our algorithm as outlined in the previous section, primarily consists of three steps: initial BFS/SSSP (step 1), iterative refinement (steps 2-5), and relaxation (step 6). We follow a {\em common initial-relax algorithm} for both the initial BFS and relaxation steps, passing a set of {\em active} vertices as input to the common algorithm.
In the initial BFS/SSSP step, we invoke the algorithm passing the source vertex as the active vertex and essentially follow the frontier-based algorithm for the forward phase outlined earlier in Figure \ref{seq-BC}. In this step, the $\sigma$ values of the vertices are computed as shown in the frontier-based algorithm.

The iterative refinement consists of two kinds of updates to the distances of the nodes: 1. updating the distances of border nodes of a partition using the distances of border nodes in the other partition using the weights of edge cuts (steps 2 and 4), and 2. updating the distances of border nodes in a partition using the distances of border nodes in the same partition, using the border matrix of the partition (steps 3 and 5). During the first kind of updates, the $\sigma$ values of the border nodes are updated similar to the calculations shown in the frontier-based algorithm of Figure \ref{seq-BC}. For the second kind of updates, during the border matrix computations of distances (section IV.B), we also compute a $\sigma$ matrix, $\sigma M$ where $\sigma M[i,j]$ denotes the $\sigma$ value of a border node $j$ in the BFS/SSSP computations with border node $i$ as the source vertex. Thus, during the second kind of updates of distance of the border node $j$ due to border node $i$ with the current source $s$, we update the $\sigma$ value of $j$ as $\sigma[j] = \sigma[i] + \sigma M[i,j]$, where $\sigma[i]$ is the value for border node $i$ computed using the current source vertex $s$. Thus, at the end of the step, all the border nodes will obtain their correct $\sigma$ values.

In the relaxation step for a partition, we invoke the common {\em initial-relax algorithm}, passing as input the set of border nodes of the partition as the active vertices. However, some of the distances of the border nodes can be smaller than the distances of some nodes in the partition found in the initial BFS/SSSP step. This can lead to a situation of updating the distances of some of the nodes to smaller values in the relaxation step. In such cases, the earlier contribution to a $\sigma$ value of a node $v$ due to one of its predecessor nodes, $u$, will have to be nullified. To achieve this, we maintain $edge\sigma(u,v)$ of the edge $(u,v)$ as the $\sigma$ value of the node $u$. When the $\sigma$ value of the node $v$ is updated in the relaxation step, the earlier contribution due to the predecessor $u$, $edge\sigma(u,v)$, is subtracted from the $\sigma$ value of $v$. The use of $edge\sigma$ values for correct relaxation is based on the approach followed in Prountzos and Pingali \cite{prountzos-BC-ppopp2013}.

\subsection{Asynchronous and Hybrid Backward Phase}

At the end of the forward phase, the dependent nodes and edges that constitute the shortest paths, i.e., the dependent DAG, can be distributed across both the CPUs and GPUs. We have designed a novel and asynchronous hybrid CPU-GPU backward phase algorithm that minimizes the amount of CPU-GPU communications.

In our hybrid algorithm for the complete BC computations, one of the CPU threads handles the invocations of the GPU kernel, passing the input to and processing the outputs from the kernel. We denote this CPU thread as GPU handler thread. During the backward phase, the GPU handler thread invokes the GPU backward phase kernel for each distance level starting from the maximum distance of the partition in the GPU. Similarly, the other CPU threads perform the CPU backward phase starting from the maximum distance of the partition in the CPU. After each GPU kernel invocation for a level, the GPU handler thread reads a boolean variable $borderNodeinLevel$ that indicates if the current level in GPU has a border node, $b_i$, in the GPU partition whose predecessor is a border node, $b_j$, in the CPU partition. If $borderNodeinLevel$ is set to true by the GPU kernel, then the GPU handler thread reads the $\delta$ and $\sigma$ values of $b_j$. 

The backward phase computations proceed independently in both the CPU and GPU devices until a device, $d1$, reaches a level $l$ that has a border node, $b_i(l)$, in its partition having one of its children as a border node,  $b_j(l+1)$, at level $l+1$ in the partition of the other device, $d2$. In this case, the computation of $\delta[b_i(l)]$ in $d1$ requires $\sigma[b_j(l+1)]$ and $\delta[b_j(l+1)]$ from the other device, $d2$. The backward phase computations in $d1$ wait till the device $d2$ reaches and finishes computations for level $l+1$. Thus, the number of true CPU-GPU synchronizations and communications (discounting the synchronizations due to kernel invocations by the GPU handler thread) is related to and limited by the number of border nodes in the CPU partition, unlike the BSP model of the earlier hybrid strategy in Totem \cite{gharaibeh-oxenChicken-pact2012} in which both the CPU and the GPU wait for each other to complete each level and where CPU-GPU communications of boundary data structures are performed every level.


\subsection{Practical Implementation}
\label{practical_imp}

We leverage the optimizations in the existing literature along with our own novel techniques.
Our CPU BFS/SSSP and relaxation computations are based on the frontier-based vertex-parallel algorithm of Madduri et al. \cite{madduri-multithreadedBC-ipdps2009}. We created OpenMP threads equal to the number of CPU cores, and used one of the threads as GPU handler thread and the other threads for performing the BC computations on the GPU. Our GPU implementation of these steps is based on our extension to the frontier-based edge-parallel BFS code of LoneStar-GPU version 2.0 \cite{lonestargpu-web}, the latest version at the time of writing. For partitioning, we use METIS \cite{karypis1998multilevelk, karypis-metis-siamscicom1998} which gives partitions of equal sizes with minimal edge cuts.


\subsection{Variable Partitioning and Backward Phase Optimizations}
\label{back_opt}

METIS partitions the graph into equal partitions. Equal sized partitioning for a heterogeneous architecture such as a multi-core CPU and GPU will result in inefficiency and poor utilization due to the different performance in the two devices for the computations. Hence, we split the computations in the ratio of performance on the CPU and GPU. To obtain the ratio, we initially partition the graph into partitions of equal size, one each for CPU and GPU, and then execute the BFS/SSSP calculations on the two devices with their respective partitions. The run-times on CPU and GPU are recorded, and CPU-GPU performance ratio is calculated using the reciprocals of the runtimes. The ratios were obtained using ten source vertices, each for CPU and GPU, for BFS/SSSP and average of the ratios is obtained.  We chose BFS/SSSP kernel as it corresponds with the BC application dealt with in this work. We used PATOH \cite{catalyurek2011} for obtaining the variable partitioning. 

In the case of the backward phase on both CPU and GPU, we use topology-driven \cite{nasre2013data} parallelization method. We experimented with both vertex based and edge based parallelization for the backward phase. The vertex based implementation uses a {\em pull} mechanism at a given level to obtain the $\delta$ and $\sigma$ values from its successors. The edges of each vertex are handled by a single thread. In the edge based implementation, the edges between the vertices at a given level and the next are distributed to the threads such that each thread processes a set of edges with one edge per thread in most cases. A thread then uses a {\em push} mechanism to modify the $\delta$ values of the predecessor vertex. Vertex-based parallelization provides the advantage that it avoids locking to compute the values for a vertex, unlike edge-based implementation which requires locking for simultaneous updates of a vertex by multiple threads processing different edges of the vertex. However, the advantage of edge-based parallelization is that it supports larger amounts of parallelism since the number of edges is greater than the number of vertices.

We found that vertex-based parallelization provided better performance for the CPU backward phase algorithm due to the limited number of threads and high cost of locking in the CPU. In contrast, edge-based parallelization provided better performance on the GPU due to the large amount of threads and parallelism available on the GPU. Hence, we adopted vertex-based parallelization on the CPU and edge-based parallelization on the GPU for backward phase.
For the GPU implementation, we used CUB \cite{merrill_high_2011} library primitives for high performance atomic constructs for locking.

%% file: exp_res.tex
\section{Experiments and Results}
\label{exp_res}

All our experiments were performed on a GPU server consisting of a dual octo-core Intel Xeon E5-2670 2.6 GHz server with CentOS 6.4, 128 GB RAM, and 1 TB GB hard disk. The CPU is connected to two NVIDIA Kepler K20 cards. We denote our hybrid strategy as HyBIR ({\bf H}ybrid {B}C using {\bf I}terative {\bf R}efinement). We compared our HybBIR code with the Totem hybrid code, and also with the standalone CPU code base used in our hybrid algorithm. We ran the CPU portions of our HyBIR, Totem and the CPU-standalone codes with 16 OpenMP threads running on the 16 CPU cores, and ran the GPU portions of our HyBIR and Totem with block size of 1024 threads. HyBIR used 18 blocks, while Totem dynamically varied the number of blocks throughout the execution. We also compared with the state-of-art GPU implementation by Mclaughlin et al. \cite{mclaughlin-scalableBCGPU-sc2014} using similar configurations.

The graphs used in our experiments are shown in Table \ref{graph-specs}. The directed graphs were converted to undirected versions. The graphs were taken from the 10th DIMACS challenge \cite{dimacs9-web, dimacs10-web}, the University of Florida Sparse Matrix Collection \cite{florida-web}, and the Laboratory for Web Algorithmics \cite{BMSB}. The table also gives the approximate diameters of the graphs based on the maximum distances in the shortest paths found in our experiments. As shown, the graphs belong to different categories and have different characteristics. Power law graphs such as uk-2014 have large maximum degrees, whereas graphs like road networks have uniform degree distribution. The former graphs are also small world graphs having smaller diameters whereas the latter have larger diameters (except Europe and Germany graphs).

\begin{table}
 \footnotesize
 \centering
 \begin{tabular}{||p{0.7in}|p{0.29in}|p{0.32in}|p{0.28in}|p{0.25in}|p{0.25in}|p{0.25in}||}
  \hline\hline
  {\bf Graph} & {\bf $|V|$ ($10^6$)} & {\bf $|E|$ ($10^6$)} & {\bf Approx. Diam.} & {\bf Avg. Deg.} & {\bf Max. Deg.} & {\bf \# src} \\
  \hline\hline
  \multicolumn{7}{||l||}{Road networks} \\
  \hline
   USA-Full & 23.95 & 57.71 & 6261 & 2.41 & 9 & 100 \\
   USA-CTR & 14.08 & 34.29 & 3826 & 2.44 & 9 & 100 \\
   Europe & 50.91 & 57.20 & 206 & 1.12 & 24 & 100 \\
   Germany & 11.55 & 13.19 & 117 & 1.14 & 21 & 1000 \\
   \hline
   \multicolumn{7}{||l||}{Delaunay Networks} \\
   \hline
   delaunay\_n24 & 16.78 & 83.89 & 1313 & 5.0 & 112 & 100 \\
   delaunay\_n25 & 33.55 & 167.78 & 1857 & 5.0 & 124 & 100 \\
   \hline
   \multicolumn{7}{||l||}{Social Networks} \\
   \hline
    uk-2014-host & 4.7 & 50.82 & 110 & 10.65 & 98k & 1000 \\
    web-edu & 9.85 & 55.31 & 221 & 5.62 & 3841 & 1000 \\
    \hline
   \multicolumn{7}{||l||}{Synthetic graphs} \\
    \hline
   nlpkkt200 & 16.24 & 415.75 & 78 & 25.60 & 180 & 1000 \\
   nlpkkt240 & 27.99 & 718.49 & 114 & 25.67 & 180 & 100 \\
   rgg\_n25 & 33.55 & 324.65 & 1256 & 9.67 & 61 & 100 \\
   \hline
 \end{tabular}
\caption{\small{Graphs for Experiments. USA graphs \cite{dimacs9-web}, Europe and Germany \cite{dimacs10-web}, delaunay graphs \cite{dimacs10-web}, social networks \cite{BMSB}, nlpkkt graphs \cite{florida-web}, and random geometric graph \cite{dimacs10-web}}}
\label{graph-specs}
\end{table}

For a given graph, we execute the methods for $k$ random sources, where $k$ was set to $1000$ or $100$, depending on the time consumed for the graph. The last column of the table shows the number of sources for which the BC computations were performed. We primarily show results in terms of TEPS (traversed edges per second), measured as $\frac{m \times k}{t}$ where $m$ is the number of edges of the graph, $k$ is the number of sources for BC computations, and $t$ is the time taken. In cases where the individual stages of the algorithms are analyzed, we report the execution times.

\subsection{Comparison with Totem}

We first compare the total times taken by the Totem hybrid code and our HyBIR approach for million source nodes. We obtain the times by executing each graph for the number of sources mentioned in the last column of Table \ref{graph-specs}, and extrapolating the time to million sources. For these experiments, we used the basic implementation of our HyBIR algorithm without the variable partitioning and the backward phase optimizations discussed in Section \ref{back_opt}. This is to primarily compare the algorithmic models of our independent computations and iterative refinement strategies with the Totem's BSP approach. We use 50-50 equi-partitioning of the graphs in both the Totem and our HyBIR models. In our model, the equal partitioning is achieved by METIS. 

\begin {figure}
\centering
\subfigure[Road Networks] {
\includegraphics[scale=0.3]{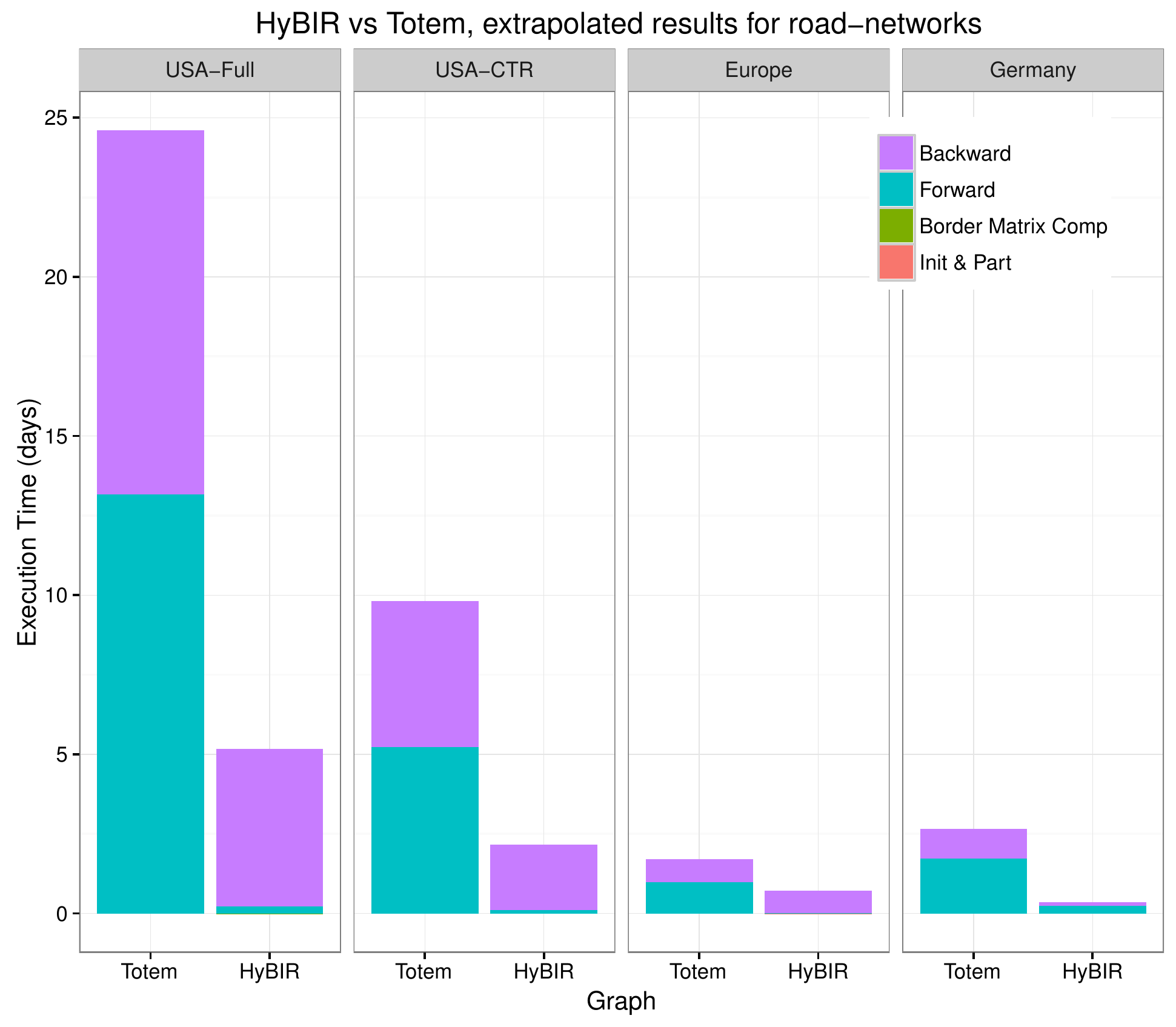}
\label{extrapolations_equal_roadnetwork}
}
\subfigure[Delaunay, Social Network and Synthetic Graphs] {
\includegraphics[scale=0.3]{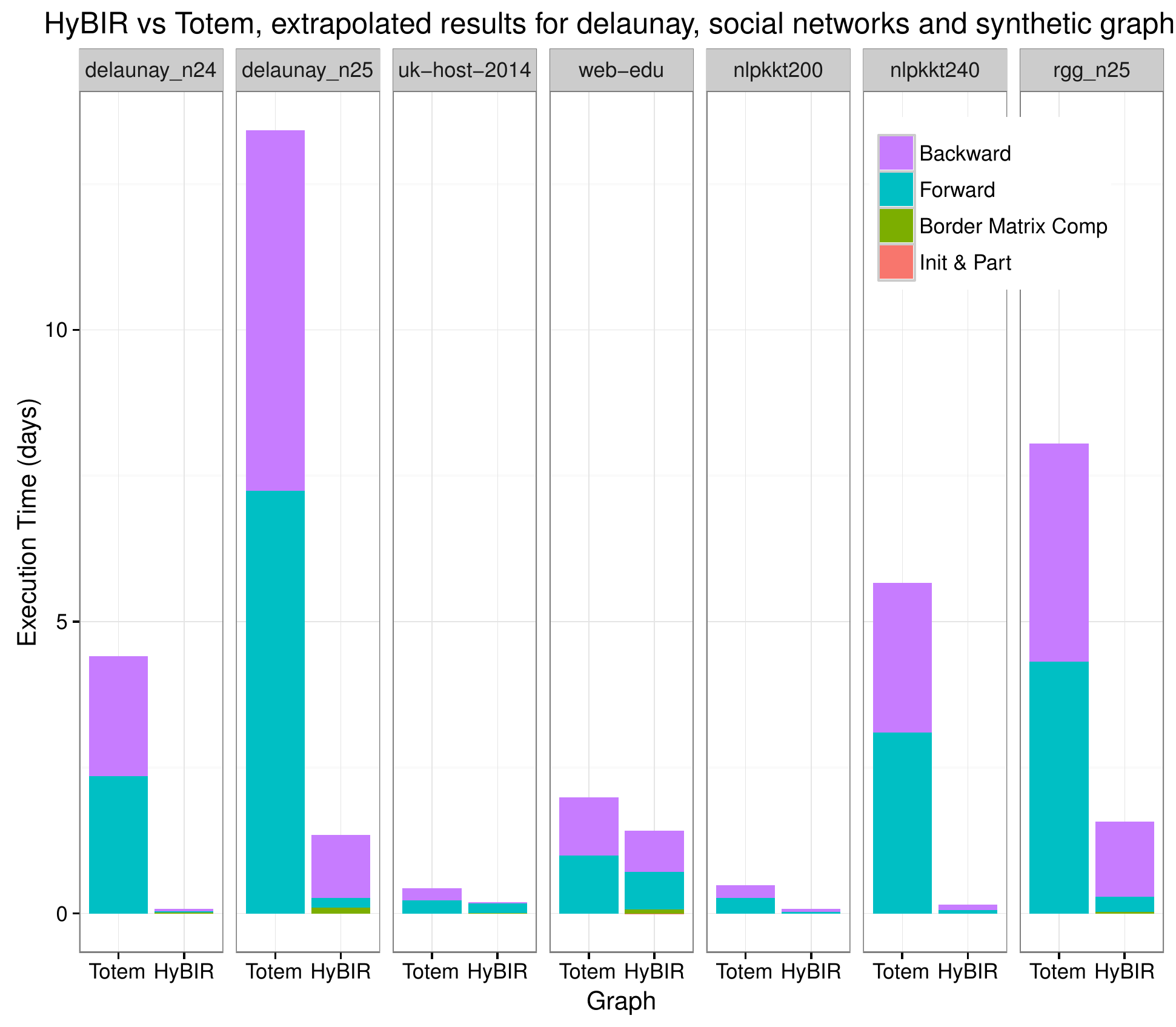}
\label{extrapolations_equal_delsocsyn}
}
\caption{\small{Totem vs HyBIR, component analysis for extrapolated for $10^6$ sources}}
\label{extrapolations_equal}
\end{figure}

Figure \ref{extrapolations_equal} shows the results including the overheads for both HyBIR and TOTEM. We find that HyBIR gives 29-98.2\%, with an average of 77.07\%, reduction in execution time when compared to Totem. The performance improvement in the forward phase is 24-99\% with an average of 83\%, while the performance improvement in the backward phase is 25-97\%, with an average of 72\%. The superior performance improvement in the forward phase is mostly due to independent computations on the partitions, unlike the BSP approach in Totem. We also find that partitioning, border matrix computations and initializations consume negligible times with respect to the BC computations. This demonstrates that our algorithm efficiently harnesses the multi-source property of BC computations, in which one-time partitioning and border matrix computations are used for multiple source nodes.

\begin {figure}
\centering
\includegraphics[scale=0.28]{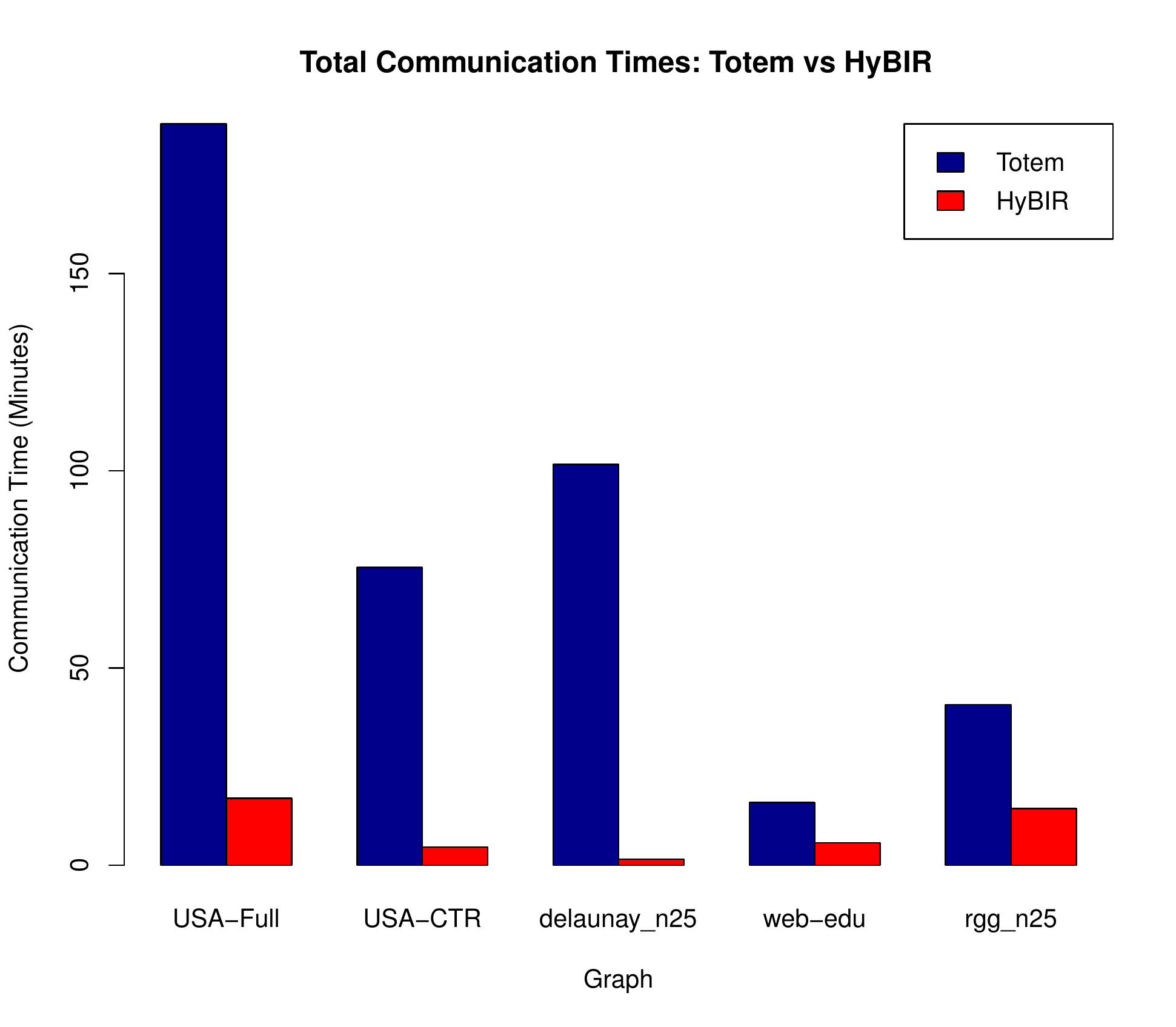}
\caption{\small{Total Communication Times}}
\label{totalcommtimes}
\end{figure}

ToTem and other approaches that perform graph computations across multiple resources follow level-synchronous BSP approach that involve coordination and communication across the resources for each level. Our HyBIR approach mostly performs independent computations on the CPU and GPU resources. To verify, we compared the total CPU-GPU communication times in Totem and HyBIR. Figure \ref{totalcommtimes} shows the communication times for the sources shown in Table \ref{graph-specs}. We find that our HyBIR approach performs 64-98.5\% less communications than the BSP approach of Totem.

We also compared our novel backward phase algorithm with Totem's backward phase.
In most cases, the number of synchronizations and communications in our approach is less than 5, and is independent of the size of the graph. The number of synchronizations and communications in Totem is equal to the maximum distance of the shortest paths from the source or the maximum number of levels found in the forward phase and ranges from 74 to 6093 for our graphs. In our backward phase algorithm, the number is limited by the number of border nodes. Thus, our hybrid approach leverages the default property of  existing partitioning tools that attempt to minimize edge cuts and the number of border nodes, which in turn result in minimum CPU-GPU synchronizations in our approach.

\subsection{Variable Partitioning and Backward Phase Optimizations}

We show the effects of our optimizations, primarily, variable partitioning and backward phase optimizations discussed in Section \ref{back_opt}. We also experimented with the variable partitioning in Totem. While variable partitioning reduced the Totem execution times by about half in most of the graphs, the times were still 2.5-4X higher than our HyBIR's base implementation results shown earlier in Figure \ref{extrapolations_equal}. Hence, in this section, we show the comparisons only between the optimized and base implementations of HyBIR.

Table \ref{utilization} shows the  utilization percentages on both CPU and GPU for the equal and variable partitioning implementations on some of the graphs. We find that variable partitioning provides almost equal utilization on each CPU and GPU due to the proportional workloads provided to each processor. The table shows that the variable partitioning significantly improves the CPU-GPU utilization.

\begin{table}
 \footnotesize
 \centering
 \begin{tabular}{||p{0.7in}|p{0.35in}|p{0.35in}|p{0.35in}|p{0.35in}||}
  \hline\hline
  Graph & \multicolumn{2}{|c|}{Equal partitioning}  &  \multicolumn{2}{|c|}{Variable partitioning} \\ \hline
       & CPU Util. (\%) & GPU Util. (\%) & CPU Util. (\%) & GPU Util. (\%) \\
  \hline\hline
  Europe & 100 & 24 & 100 & 91 \\ 
  delaunay\_24 & 100 & 26 & 89 & 100 \\ 
  uk-2014-host & 100 & 29 & 86 & 100 \\
  web-edu & 100 & 35 & 100 & 89 \\
  delaunay\_n25 & 23 & 100 & 95 & 100 \\
  \hline\hline
\end{tabular}
\caption{\small{Equal and Variable Partitioning processor utilizations}}
\label{utilization}
\end{table}

\begin {figure}
\centering
\includegraphics[scale=0.32]{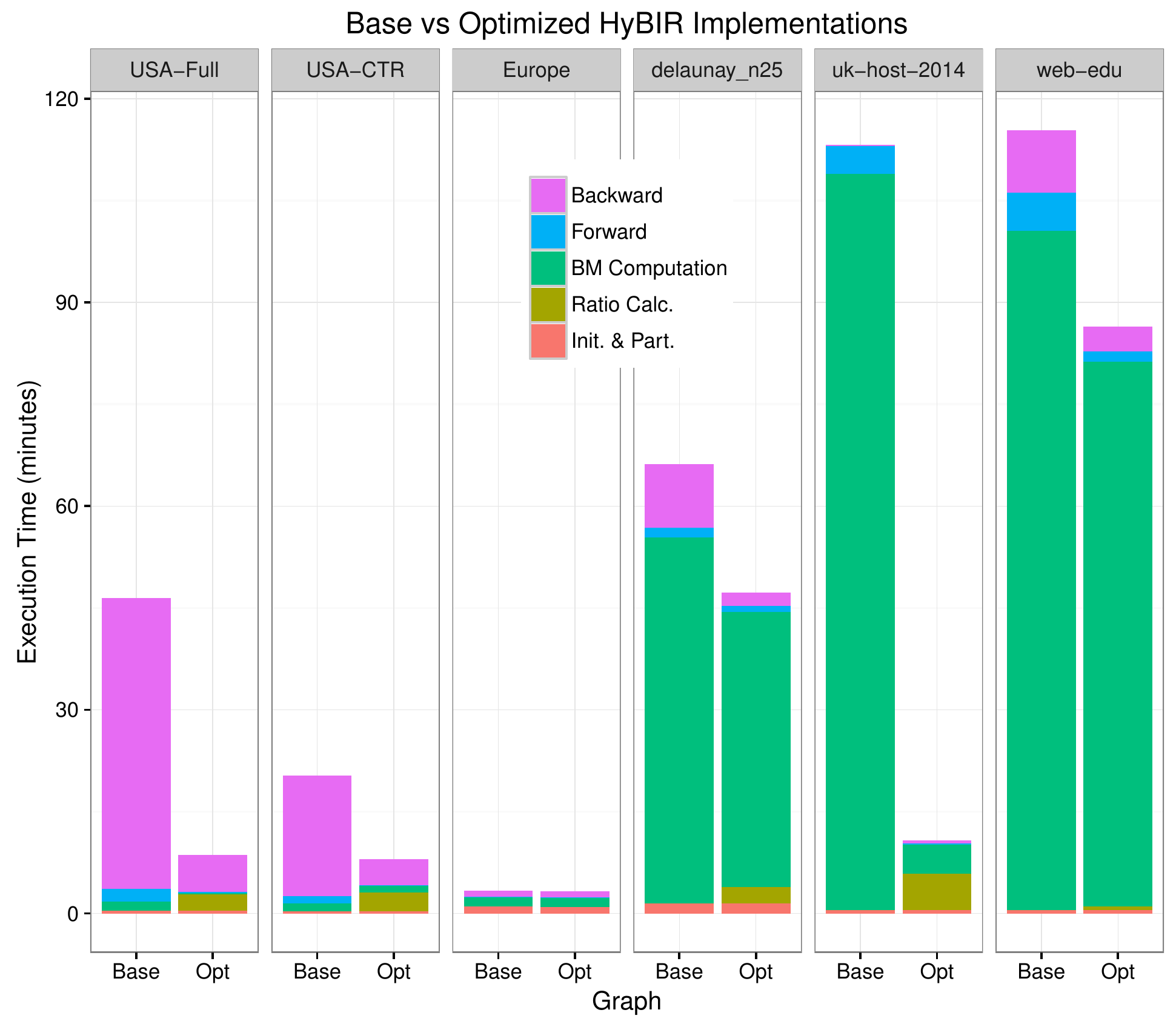}
\caption{\small{Base vs Optimized Implementations of HyBIR in terms of execution times}}
\label{actual_basevsopt}
\end{figure}

Figure \ref{actual_basevsopt} compares the base and optimized implementations of HyBIR in terms of execution times for the number of sources shown in Table \ref{graph-specs}, for some of the graphs. The variable partitioning technique performs vastly superior in comparison to equal partitioning, achieving speed ups of upto 10x, with an average speedup of around 3x. The improvements in forward phase is due to variable partitioning, while the improvements in backward phase are due to both variable partitioning and backward phase optimizations. In cases, where the forward phase timings are about equal, thus implying 50-50 performance ratio between CPU and GPU, the improvements are due to backward phase optimizations. The figure also shows that the road network graphs have much less overheads than the social network and synthetic graphs,  hence we can clearly see the performance improvements in the forward and backward phase of the BC algorithm for the road network graphs. For the social network and synthetic graphs, the border matrix computations consume most of the times. However, these times are amortized in calculations for large number of sources as shown earlier in our extrapolation results.

\subsection{Comparison with CPU Standalone Code}

\begin {figure}
\centering
\subfigure[Road Networks and Delaunay Graph]{
\includegraphics[scale=0.21]{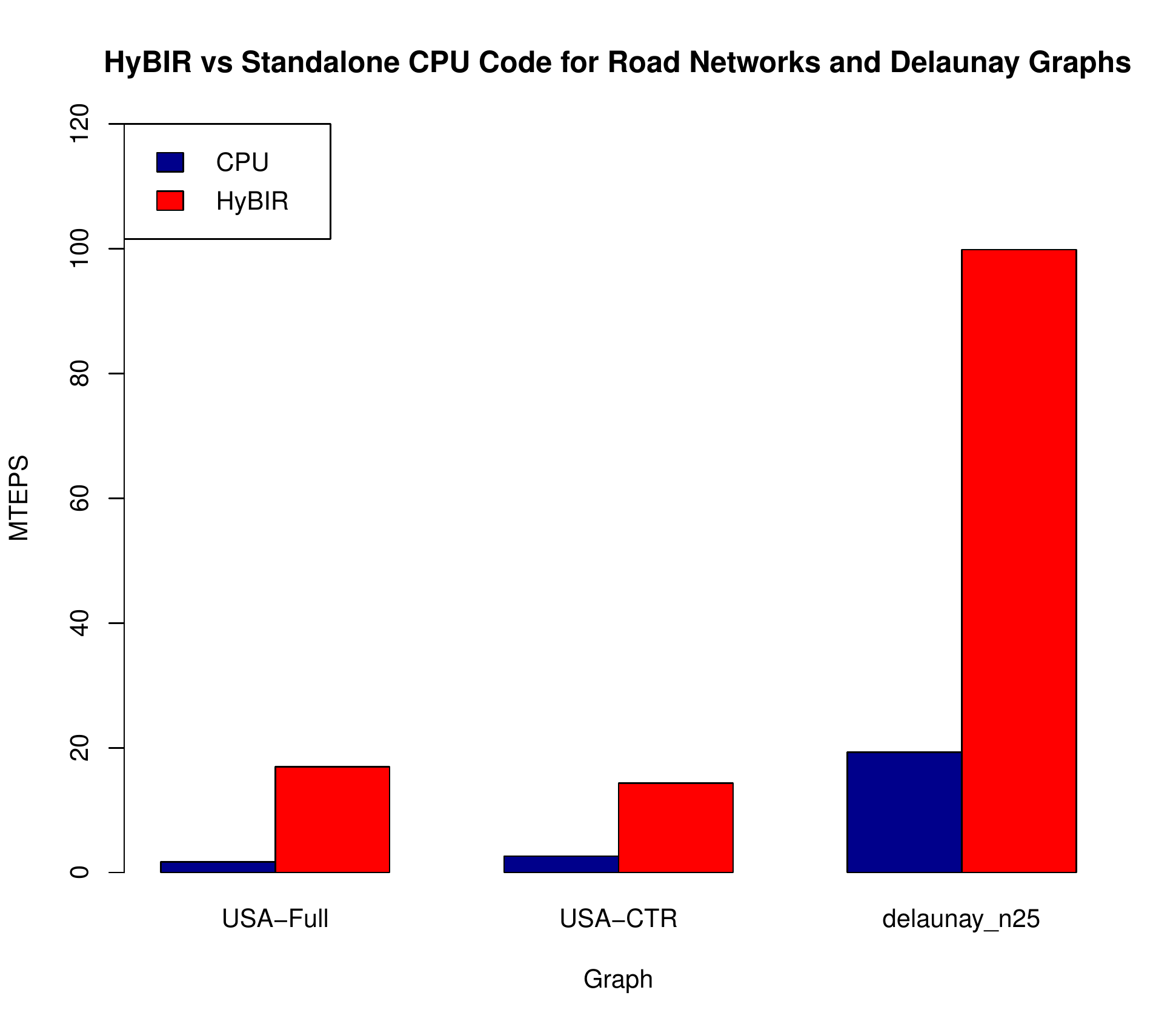}
\label{hybridvsstandalone_roadnetwork}
}
\subfigure[Social Network and Synthetic Graphs] {
\includegraphics[scale=0.21]{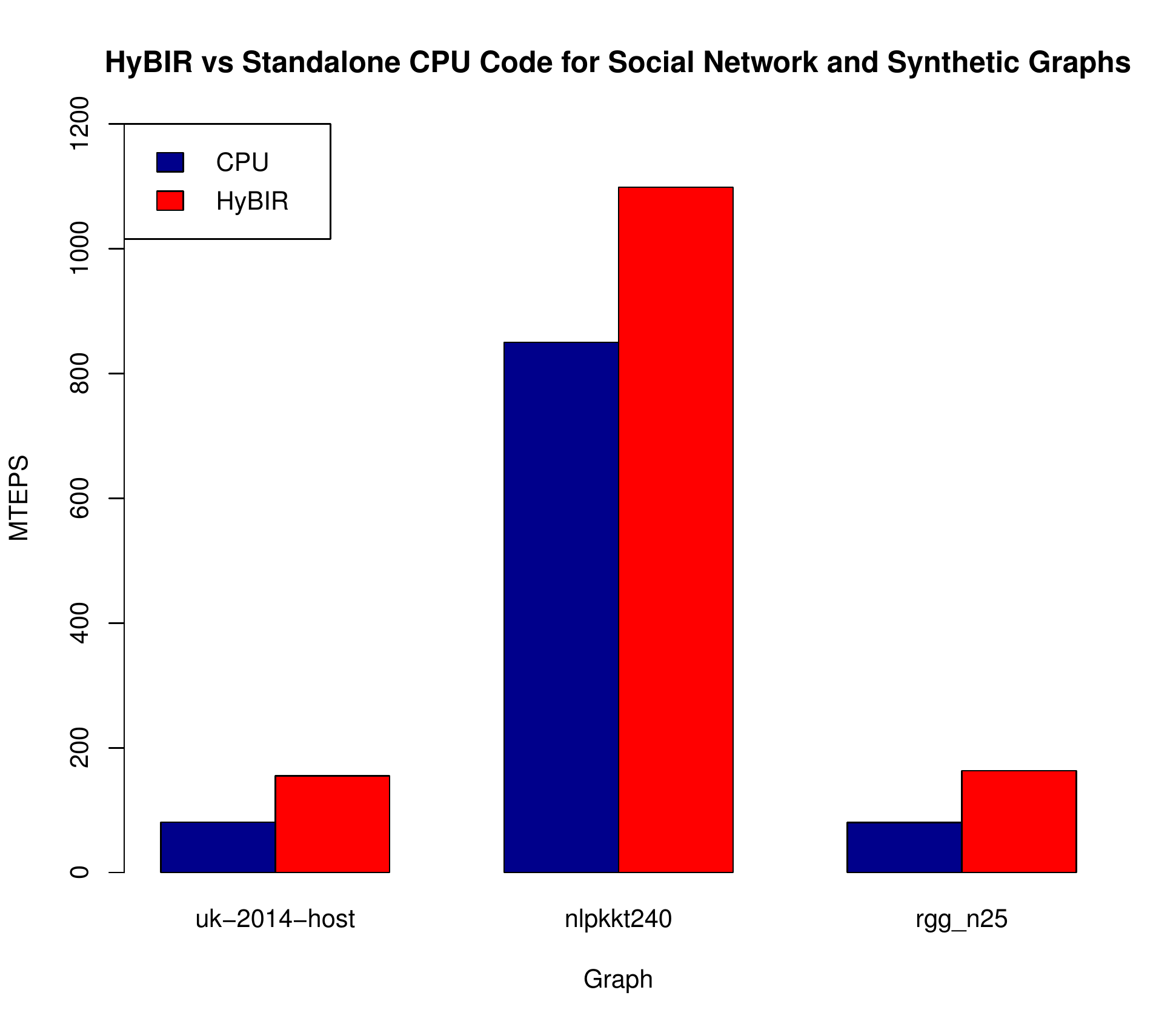}
\label{hybridvsstandalone_delsocsyn}
}
\caption{\small{HyBIR vs CPU standalone code}}
\label{hybridvsstandalone}
\end{figure}

We compare HyBIR with the CPU standalone implementation. The initial BFS/SSSP step of our hybrid algorithm can be pipelined, implying that when one of the CPU and GPU resources execute this step for one source, the other resource can perform look-ahead computations for the next source. However, such pipelining is not possible for the CPU standalone implementation. Figure \ref{hybridvsstandalone} shows the comparisons in terms of MTEPS for some of the graphs. Similar trends were observed for the other graphs. In all cases, HyBIR gives 2-8X better performance than the CPU version since HyBIR harnesses the massive amount of parallelism from the GPU along with the minimal communications and synchronizations between the CPU and GPU partitions. The results demonstrate that hybrid implementations can make use of the extra power due to the GPUs to improve the performance of the CPU-only implementations.

\subsection{Comparing with a state-or-art GPU Implementation}

Finally, we compare HyBIR with state of art many-core GPU standalone implementation by Mclaughlin et al. \cite{mclaughlin-scalableBCGPU-sc2014}. We extrapolate the execution times for all the sources in each graph, and also included the initialization, partitioning, ratio calculation and border matrix computation overheads. The results are shown in Table \ref{HyBIR-kamesh-adam}.

\begin{table}
\centering
\footnotesize
\begin{tabular}{|p{0.75in}|p{1.0in}|p{1.0in}|}
\hline 
Graph  & McLaughlin et al. (in Hours) & HyBIR (in Hours) \\  
\hline 
uk-host-2014 & 963.14 & 464.22 \\
\hline 
web-edu & 1046.13 & 575.97 \\
\hline 
Other 9 graphs & error & 234.83 ~--~ 22470.59 \\
\hline 
\end{tabular}
\caption{\small{HyBIR comparison with Mclaughlin et al.'s work}}
\label{HyBIR-kamesh-adam}
\end{table}

As shown in Table \ref{HyBIR-kamesh-adam}, the GPU implementation by Mclaughlin et al. was not able to accommodate and execute nine of the eleven graphs. Their implementation performs coarse-grain parallelization of a batch of sources at a time. Each source is executed by a single {\em SM} of the GPU in parallel. This severely limits the graph sizes that can be accommodated. Hence their approach was not able to execute these graphs due to the compounding memory requirements for the batch of sources executing at once. For the remaining two graphs, namely {\em uk-host-2014} and {\em web-edu}, HyBIR performs about 2X better than McLaughlin et al.'s implementation. In their approach, the parallelization of each source is limited by the small number of threads available per SM of the GPU. Hence, the performance of a batch of sources is limited by the worst performing source in the batch. This effect is predominant for these two large graphs.

%% file: confut.tex
\section{Conclusions and Future Work}
\label{con_fut}

In this work, we had developed a novel fine-grained CPU-GPU hybrid betweenness centrality (BC) algorithm that partitions the graph and performs independent computations on the CPU and GPU. 
We have also designed a novel backward phase algorithm that performs as much independent traversals on the CPU and GPU as possible. Our evaluations show that our hybrid approach gives 80\% improved performance over an existing hybrid strategy that uses the popular BSP approach. Our hybrid algorithm also gives better performance than the CPU-only version, and can explore graphs that cannot be accommodated in the GPU memory. In future, we plan to explore dynamic partitioning strategies based on dynamic CPU-GPU performance ratios, and extend our algorithm for multiple partitions to utilize a large number of CPU and GPU resources in tandem for exploring big-data graphs.